\documentclass[reprint,amsmath,amssymb,aps,nofootinbib,prd]{revtex4-2}
\usepackage{graphicx} 
\usepackage{dcolumn} 
\usepackage{bm} 
\usepackage{aas_macros} 

\begin{document}

\title{Thick Disks, Thin Hopes: Suppressed Capture and Merger Rates in AGN}

\author{Yashvardhan Tomar}
\email{yash@caltech.edu,\,phopkins@caltech.edu}
\author{Philip F. Hopkins}
\affiliation{Cahill Center for Astronomy and Astrophysics, California Institute of Technology, Pasadena, CA 91125, USA}
\author{Kyle Kremer}
\affiliation{Department of Astronomy \& Astrophysics, University of California, San Diego; La Jolla, CA 92093, USA}

\date{\today}

\begin{abstract}
Multiple models have been suggested over the years to explain the structure and support of accretion disks around supermassive black holes -- from the standard thin thermal-pressure-dominated $\alpha$-disk model to more recent models that describe geometrically thicker radiation or magnetic or turbulence-dominated disks. In any case, objects embedded in the disk (e.g.\ compact objects, stars, gas, dust) can undergo gravitational and hydrodynamic interactions with each other leading to interesting processes such as binary interaction/capture, gravitational wave merger events, dynamical friction, accretion, gap opening, etc. It has long been argued that disks of active galactic nuclei (AGN) can enhance the rates for many of these events; however, almost all of that analysis has assumed specific thin-disk models (with aspect ratios $H/R \lesssim 0.01$). We show here that the rates for processes such as these that are mediated by gravitational cross-sections has a very strong inverse dependence on the thickness $H/R$ (scaling as steeply as $(H/R)^{-8}$), and $H/R$ can vary in the outer disk (where these processes are often invoked) by factors $\gtrsim 1000$ depending on the assumed source of pressure support in the disk. This predicts rates that can be lower by tens of orders-of-magnitude in some models, demonstrating that it is critical to account for disk parameters such as aspect ratio and different sources of disk pressure when computing any meaningful predictions for these rates. For instance, if magnetic pressure is important in the outer disk, as suggested in recent work, capture rates would be suppressed by factors $\sim 10^{10}-10^{20}$ compared to previous studies where magnetic pressure was ignored.
\end{abstract}

\maketitle

\section{Introduction}\label{sec:intro}

Active galactic nuclei (AGN) and quasars are powered by accretion disks around a central supermassive black hole (SMBH), extending from near-horizon (the gravitational radius $R_{\rm S} = 2 G M /c^{2}$, of order au) to scales around 0.1-10 pc \citep{DiskSize}, and supplying mass to the SMBH at rates as high as $\sim 10 \,{\rm M}_\odot$ yr$^{-1}$ \citep{DiskMassFlowRate} thereby powering highly luminous emission.
While this is well-established, basic physical properties of the disk, including its thickness/aspect ratio $H/R$, and the dominant physical form of the energy density or pressure within the disk (supporting its vertical structure and competing against self-gravity within the midplane) remain deeply uncertain, especially in the outer disk (radii $\gg 100\,R_{\rm S}$). 
The classic $\alpha$-disk model of \citet{SS73,novikov.thorne:1973.astro.of.bhs} assumes the dominant pressure is thermal, but alternative models have been proposed where radiation \citep{paczynsky.wiita:1980.slim.disk,abramowicz:1988.slim.disks} or supersonic/Alfvenic turbulence \citep{SG2003,tqm} or strong magnetic fields \citep{Pariev,BP2007,johansen.levin:2008.high.mdot.magnetized.disks,gaburov:2012.public.moving.mesh.code} instead dominate the disk pressure. These can have qualitatively large effects on the disk structure (including its aspect ratio, density, surface density, accretion timescales, etc.).

Galactic nuclei also present a dense environment of stars and compact objects \citep{gcd1,gcd2,gcd3}, and interactions within the disk (for objects formed or gravitationally captured into the accretion disk) can significantly affect the dynamical evolution of these objects. AGN disks can act as sites of gaseous accretion onto stars or compact objects, as well as inducing repeated mergers by driving migration via dynamical friction and Type I/II torques, reducing the inclination of intersecting orbits (leading to disk capture, and increasing the number of objects embedded in the disk), and enhancing the formation and hardening of binaries orbiting the SMBH. This makes them interesting locations for numerous interesting astrophysical phenomena. A few examples which have been discussed extensively in the literature include:

\begin{itemize}
    \item These have been proposed as important channels for growth of intermediate mass black holes (IMBHs). As IMBH seeds migrate within the disk, they might grow at super-Eddington rates, both via gaseous accretion as well as collisions with stars and compact objects at low relative velocities \citep{imbh1}. These IMBHs, if sufficiently massive, can open gaps in the AGN disks which can cause dips in the spectral energy distribution (SED) of the disk \citep{imbh2}. Accretion onto IMBHs that do not open gaps can produce soft X-ray excess relative to the continuum emission. Gravitation waves (GW) emitted during IMBH inspiral and collisions can be potentially detected by LISA \citep{LISA}.

    \item Stellar mass BH binaries can get captured in AGN disks as they migrate and encounter each other at small relative velocities, much smaller than the encounter velocities in gas-free nuclear star clusters \citep{Leigh2018,imbh1}. They can harden via gas dynamical friction, and Type I/II torques and then merge (see \cite{EvolutionOfCompactObjsInDisks} and the references therein for a detailed description of how these different processes operate). These can, therefore, be important sources of gravitational waves for LIGO/Virgo/KAGRA. Further, the inspiral and merger of the binary BH within the gaseous medium can sustain luminous electromagentic emission due to super-Eddington accretion \citep{Stone2017} and from shocks produced by orbital motions of the binary \citep{Farris2012}. A number of studies have tried to constrain the binary BH merger rate for LIGO from AGN disks.

    \item Besides binary BH mergers, there can also be mergers between other compact objects such as neutron stars (NS) and white dwarfs (WD). In addition to GW signatures, muffled electromagnetic signatures arising from events in the disk like tidal disruptions of NS by BH, kilonovae from NS–NS mergers, and supernovae (SNe) from WD mergers or tidal disruptions might be detectable in large surveys of AGN \citep{AllCompacts}.

    \item \cite{LongStar1,LongStar2} have proposed the existence of super-massive stars forming in AGN accretion disks, seeded at large radii beyond the gravitational radius of influence of the SMBH -- beyond which the disk becomes self-gravitating and prone to fragmentation -- and then migrating inward, possibly also opening a gap in the disk, and leading to gravitational wave signals (potentially detectable in the LISA band) as they inspiral and merge with the central SMBH. Further, \cite{Jermyn_2022} have proposed the existence of ``immortal'' massive stars in AGN disks. Because the timescale at which the accretion rate in the AGN disk can replenish the hydrogen mass in the inner disk stars can be much lower than the timescale at which nuclear burning and mixing processes occur in them, these stars can be extremely long-lived. Outflows from these stars can also potentially help resupply gas in the disk, thereby helping extend the disk lifetime.
\end{itemize}

The dynamical processes at play above, like binary capture via gravitational encounters, dynamical friction, Bondi-Hoyle-Lyttleton accretion, gravitational focusing, gap-opening, Type-I/II/III migration etc.\ all proceed at a rate governed by the gravitational cross-section of the compact object/star participating in that process. In this paper (\S\ref{sec:Formalism}), we show that those cross-sections and rates are highly sensitive to the disk aspect ratio, and therefore physical models for the dominant source of pressure, in the accretion disk. This is important because almost all the historical work on these rates, including the references above, has considered a similar subset of background disk models, which generally assume a ``razor-thin'' geometry and neglect some of the pressure sources above (like magnetic fields). We show how the assumed disk physics (\S\ref{sec:models}) produces qualitatively different aspect ratios and disk profiles, which in turn produce vastly different predictions for the rates of all the events above (\S\ref{sec:rates}). We summarize the implications in \S\ref{sec:implications}.

\section{Rate Estimation}\label{sec:Formalism}

We consider an accretion disk around a central SMBH of mass $M_c$. Depending upon the physical mechanism which provides the dominant vertical pressure support, the scale-height of the disk $H(R)$ will scale with the cylindrical radius $R$ as
\begin{equation}\label{eq:scaleht}
    H(R) \simeq \frac{v_e(R)}{\Omega(R)}
\end{equation}
where, $\Omega$ is the circular orbital frequency (depending on the total enclosed mass $M_{\rm enc}(<R)$ as $\Omega \equiv \sqrt{G M_{\rm enc}/R^{3}}$), and $v_e$ is an effective velocity that can include contributions from the thermal velocity $c_s$, turbulent velocity $\delta v$, Alfv\'{e}n velocity $v_A$, and systemic velocity offsets $\Delta v$.
\begin{equation}\label{eq:coeffs}
    v_e^2 = \zeta_cc_s^2 +  \zeta_Av_A^2\sin^2{\theta} + \zeta_t\delta v_z^2 + \zeta_s\Delta v^2
\end{equation}
where $\theta$ is the angle between the magnetic field-lines and the disk-normal, $\delta v_z$ is the vertical component of the turbulent velocity (for perfectly isotropic turbulence $\delta v_z^2=\dfrac{1}{3} \delta v^2$), and the coefficients $\zeta_i$ encapsulate the relative contribution from each of the different physical processes. So, for instance, in an unmagnetised disk, $\zeta_A \rightarrow 0$; while in a low plasma-$\beta$ disk that is trans-Alfv\'{e}nically turbulent, $\zeta_t \sim \zeta_A \gg \zeta_c$ etc.

Now, given a ``test mass'' (e.g.\ BH or star or neutron star or planet, or any body that is effectively a test particle with respect to the central SMBH) of mass $m_1$, the two-body interaction rate with some target species with number density $n_t$ is 
\begin{equation}
    \Gamma(R) = n_t \sigma_1 v_\text{rel}
\end{equation}
for a process governed by the gravitational cross section $\sigma_1 = \pi b^2$ of the test mass $m_1$, where the gravitational impact parameter $b \approx \dfrac{Gm_1}{v_\text{rel}^2}$ for a target moving with velocity $v_\text{rel}$ relative to $m_1$. We argue that $v_\text{rel} \sim v_e$. This makes intuitive sense for a process where the target species is gas particles (which will have relative velocities with respect to the test mass on the scale of the thermal and turbulent speeds). Even in the case that the target species being captured are effectively collisionless stellar/BH particles, we note that since these particles (which are still effectively test particles compared to the central SMBH) either originate in the disk or are captured, they will inherit the same velocity dispersion as the gas or will settle/relax into it.\footnote{In detail they could in principle be dynamically ``hotter,'' i.e.\ have $v_{\rm rel} \ll v_{e}$ (if e.g.\ the test masses are part of a cloud/cluster simply passing through the disk), but as we will show this would radically lower their interaction rates, so only strengthens our conclusions and contributes little to the integral event rate (dominated by objects ``within the disk'', which by definition means $v_{\rm rel} \lesssim v_{e}$). But there is no mechanism operating on the timescales of interest in the disks we consider to form or maintain a dynamically much colder distribution ($v_{\rm rel} \ll v_{e}$) of test masses compared to the disk.}
The systemic offsets in the velocity $\Delta v$, become important, when say, two particles that are in eccentric orbits with varying eccentricity, inclination, and/or phase (true anomaly) interact; or say, when the test mass captures the target species along a migrating trajectory in the disk. In any case, we should stress here that the overall uncertainties in the coefficients $\zeta_i$ do not qualitatively change the results as we will proceed to show. Also, since our goal is to make a dimensional argument for the scaling relations for the capture rate, we can also drop the order-unity dependencies like $\sin{\theta}$, and replace $\delta v_z$ with $\delta v$ (or one can think that these dependencies have been subsumed into the coefficients $\zeta_i$).

Taking $n_t = \dfrac{\rho_t(R)}{m_t}=\dfrac{\Sigma_t(R)}{2H(R)m_t}$ for some surface target density distribution $\Sigma_t(R)$, we have, then,
\begin{equation}
    \Gamma(R) = f_t \frac{\Sigma(R)}{2H(R)m_t} \cdot  \pi\left(\frac{Gm_1}{v_e^2}\right)^2 \cdot v_e
\end{equation}
where we have assumed a fraction $f_t$ of the disk mass is in the target species i.e. $\Sigma_t(R) = f_t\Sigma(R)$. In principle, we expect $f_t$ to vary radially as well. But for the sake of simplicity and clarity of our argument, we do not model this dependence for now.

Relating the effective velocity $v_e$ to the vertical scale-height of the disk, and thus the aspect ratio $H(R)/R$ of the disk, as in eq. (\ref{eq:scaleht}), we can therefore express the  interaction rate for the test mass $m_1$, as
\begin{equation}\label{eq:rate}
    \Gamma(R) = \frac{\pi}{2} G^2 \frac{m_1^2}{m_t} f_t\Sigma(R) \frac{\Omega(R)^{-3}}{R^4} \left(\frac{H(R)}{R}\right)^{-4}\ .
\end{equation}

Here, a strong dependence of the gravitational interaction rate on the disk profile is immediately manifest. At a given radius in the disk, we can thus calculate the total interaction rate for all test masses $m_1$ (with number density in the disk $n_1$) within a volume d$V$, as
\begin{equation}
    {\rm d}^3\Gamma^\text{tot}(R,z) = n_1\Gamma(R) \text{ } {\rm d}V
\end{equation}
where, again we parameterize $\Sigma_1(R) = f_1 \Sigma(R)$ and neglect the radial dependence in $f_1$. Then,
\begin{equation}
    {\rm d}\Gamma^\text{tot}(R) = 2\pi \int_{z=-\infty}^{z=+\infty} \frac{\rho_1(R,z)}{m_1} \Gamma(R) R\text{ } {\rm d}R {\rm d}z
\end{equation}
which yields the total interaction rate between the primary species $m_1$ and the target species $m_t$ as a function of radius in the disk:
\begin{equation}\label{eq:netrate}
    \frac{{\rm d}\Gamma^\text{tot}(R)}{{\rm d}\ln{R}} = \pi^2 G^2 \frac{m_1}{m_t} f_1f_t \Sigma^2(R) \frac{\Omega(R)^{-3}}{R^2} \left(\frac{H(R)}{R}\right)^{-4}\ .
\end{equation}

For an accretion disk, where Reynolds-Maxwell stresses drive the angular momentum transport through the disk, we can integrate (along the azimuth $\phi$ and the disk normal direction $z$) and Reynolds-average the azimuthal momentum equation to show, in the quasi-state state (neglecting wind-loss from the disk),
\begin{equation}
    \dot{M} R^2\Omega \simeq 2\pi R^2\Sigma\left(\overline{\delta v_R \delta v_\phi} - \frac{\overline{B_RB_\phi}}{4\pi\rho}\right)
\end{equation}
where $\dot{M} = 2\pi\Sigma R\overline{v}_R$ is the mass flow rate through the disk, which is constant across the disk for quasi-steady flow. Here $\overline{(.)}$ denotes azimuthal average, and $\delta(.)$ denotes turbulent fluctuations. If the turbulence is not too highly anisotropic and $B_R$ is not much smaller than $B_\phi$ (which appears to be the case when a disk settles into a quasi-steady state even upon starting from purely toroidal magnetic field, per recent numerical simulations, see for e.g. \cite{guo2025idealizedglobalmodelsaccretion, Squire_2025}), then we can approximate,
\begin{equation}\label{eq:RMstress}
    \dot{M} R^2\Omega \simeq 2\pi R^2\Sigma\left(\delta v^2 + v_A^2\right)
\end{equation}
to argue, that for a supersonically and/or trans-Alfv\'{e}nically turbulent disk, we have, to leading order,
\begin{equation}\label{eq:mdot}
    \dot{M}\Omega \simeq 2\pi\Sigma v_e^2 \ .
\end{equation}

Here, since the thermal sound speed $c_s$ does not explicitly appear in the expression for the mass flow rate $\dot{M}$, we note that equation (\ref{eq:mdot}) gives an optimistic estimate for $\dot{M}$ in the disk (in cases where the turbulence is weak and plasma-$\beta \geq 1$); which means in our estimate, the captures rates are not limited by the accretion rate (so it goes in favour of our conclusions generally). Systemic offsets $\Delta v$ do influence the accretion rate, e.g. in eccentric disks where forcing from global lopsided/eccentric ($m=1$) modes can be drivers of accretion \citep{Hopkins_2011, Hopkins_lopsided}. Equation (\ref{eq:mdot}) can be used to recast eqns. (\ref{eq:rate}) and (\ref{eq:netrate}) into the form (again, relating the effective velocity with the aspect ratio of the disk; eqn.~\ref{eq:scaleht}),
\begin{equation}\label{eq:gamma}
    \Gamma(R) = \frac{1}{4} G^2\dot{M} \frac{m_1^2}{m_t} f_t \frac{\Omega(R)^{-4}}{R^6} \left(\frac{H(R)}{R}\right)^{-6}
\end{equation}
\begin{equation}\label{eq:effrts}
   \frac{{\rm d}\Gamma^\text{tot}(R)}{{\rm d}\ln{R}} = \frac{1}{4} G^2 \dot{M}^2\frac{m_1}{m_t} f_1f_t\frac{\Omega(R)^{-5}}{R^6} \left(\frac{H(R)}{R}\right)^{-8} \ .
\end{equation}
Note the very steep dependence $\dfrac{{\rm d}\Gamma^\text{tot}(R)}{{\rm d}\ln{R}} \propto \left(\dfrac{H(R)}{R}\right)^{-8}$. This suggests that the aspect ratio of the disk (which can vary widely between different disk models) is a crucial rate-determining factor; especially within the SMBH radius of influence where the rotation curves are dominated by the SMBH potential, meaning the disk thickness is essentially the only property of the disk that determines the capture rates. Thus, uncertainties in the disk thickness are likely to impact rate predictions much more strongly than the much less pronounced uncertainties in any other term e.g. the coefficients $\zeta_i$ in equation (\ref{eq:coeffs}).

We stress that a version of the rate equations for $\Gamma$ and $\Gamma^{\rm tot}$ in equations (\ref{eq:gamma})-(\ref{eq:effrts}) applies to essentially \textit{all} of the example processes delineated in \S\ref{sec:intro}. Obviously binary pairing, effective ``collisions'' and/or harding via three-body scattering or drag, and corresponding mergers of test masses like stars or stellar/intermediate-mass BHs within the disk are governed by the effective gravitational collision rate with some disk targets (other stars/compact objects or gas). Accretion of gas onto said test masses is rate-limited by the rate of gravitational capture of target gas particles (nuclei/atoms/molecules), i.e.\ Bondi-Hoyle-Lyttleton processes. Dynamical friction and Type-I migration rates likewise scale with the rate of strong gravitational interactions/scattering (generating the salient ``wake'' and/or resonant interactions) of a test mass with the background target density (gas or collisionless/compact objects). Gap opening/Type-II migration scales slightly differently, and is discussed separately below, but is also quite sensitive to $H/R$.

\section{Different Disk Models}\label{sec:models}

\begin{figure}
    \centering
    \includegraphics[width=0.98\columnwidth]{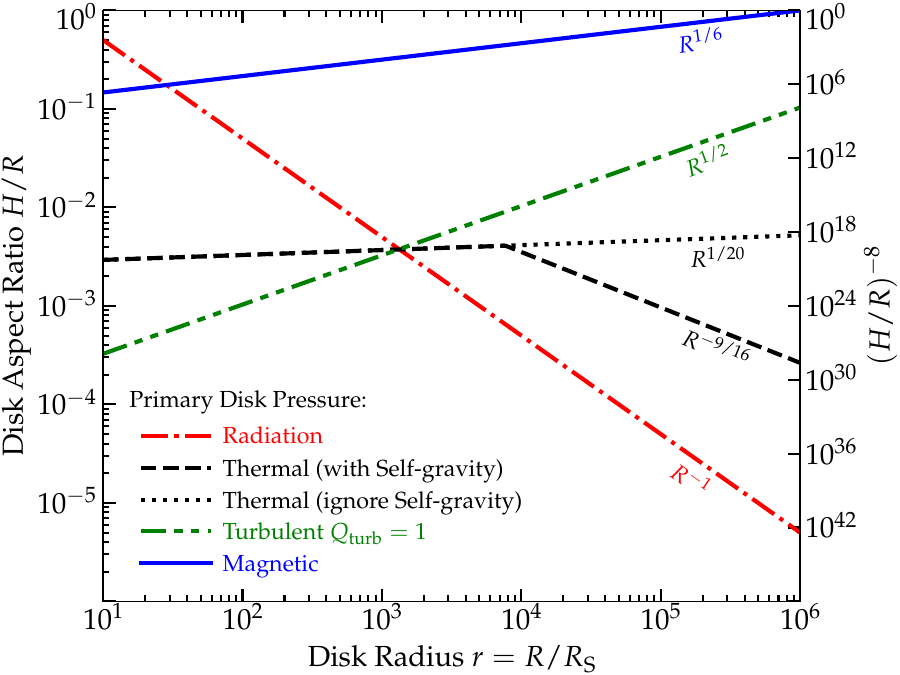}\vspace{-0.3cm}
    \caption{Disk aspect ratio or thickness $H/R$ versus radius $R$ (distance from the SMBH, in units of the Schwarzschild radius $R_{\rm S}$), for different assumptions corresponding to different assumed sources of dominant pressure within the disk (\S\ref{sec:models}). For all models we assume a central SMBH mass of $10^{8}\,{\rm M}_{\odot}$ and accretion rate $\dot{M}$ equal to half the Eddington rate $\dot{M}_{\rm Edd} \equiv L_{\rm Edd}/(0.1\,c^{2})$. For the radiation and thermal-pressure dominated disks, we assume the common literature value of $0.1$ for the $\alpha$ parameter. Depending on the assumed pressure in the disk, $H/R$ at a given radius can vary by factors $\gtrsim 1000$. The right vertical axis shows $(H/R)^{-8}$, the approximate scaling for rates of multi-body gravitational interactions (e.g.\ binary pairing, mergers, gravitational wave events, inspiral/sinking, accretion rates onto stars or stellar-mass BHs in the disk, etc.). This can differ enormously.}
    \label{fig:hor}
\end{figure}

Astrophysically, the pressure support against vertical collapse of the disk under gravity can be provided by a combination of different physical mechanisms. Here, we outline different analytic self-similar models from the literature to highlight how different dominant physics qualitatively affect the disk properties, and therefore, the two-body gravitational interaction/capture rates. For each model, we motivate simple scaling relations for the relevant disk parameters and in the subsequent section we use them to obtain the rate estimates. The radial dependence of the aspect ratio we obtain for these different disk models has been plotted in Figure \ref{fig:hor}. Explanations on how these are heuristically derived are presented in the following sections, but note the wide range in the orders of magnitude predicted by different physical models at any given radius. The effect becomes more drastic when we recall from equation (\ref{eq:effrts}) the strong dependence (aspect ratio)$^{-8}$ on the total capture rate.

\subsection{Radiation Pressure-Dominated Disks}

Many disk models have an inner region which is dominated by radiation pressure. Even in the so-called ``thin-disk'' models, these innermost regions may not necessarily be very thin, as they can have aspect ratios in excess of 0.1. The scale height is a constant, set by the mass flow rate through the disk, in the limit where the dominant opacity source is electron-scattering (as is typical for the high temperature inner regions for disks). In the case where the angular momentum transport through the disk is mediated by (effective) viscous stresses, we can write, for distances much larger than the innermost stable circular orbit (ISCO),
\begin{equation}\label{eq:visc_mdot}
    \nu\Sigma(R) = -\frac{\dot{M}}{2\pi}\left(\frac{{\rm d}\ln{\Omega}}{{\rm d}\ln{R}}\right)^{-1}
\end{equation}
where the effective viscosity is $\nu$. It is standard in many disk models to  parameterise the effective viscous stresses using an $\alpha$-prescription, $T_{R\phi} = -\alpha P$, which translates to $\nu \simeq \alpha v_eH$. Here, we use $P$ to denote the total pressure, which can in principle have contributions from radiation, thermal gas motions, disk turbulence, and magnetic fields, and note that these scalings for $\nu$ and equation (\ref{eq:visc_mdot}) apply to multiple different disk models we discuss below. 
In the specific case we consider here, where radiation pressure dominates, the disk height is largely set by the effective sonic speed (of the combined radiation-gas fluid) in the midplane,
\begin{equation}\label{eq:rad.ve}
    v_e^2\simeq c_{s,\,e}^2 \equiv {\left. \frac{P}{\rho} \right|}_{z=0} = \left(\frac{4}{3}\frac{\sigma_{SB}}{c} T_0(R)^4\right)/\left(\frac{\Sigma(R)}{2H(R)}\right) \ .
\end{equation}

In thermal radiative balance where the viscous heating rate is balanced by radiative losses from an optically-thick disk,
\begin{equation}\label{eq:heat_balance}
    Q_+ =-\nu\Sigma(R)\Omega(R)^2\left(\frac{{\rm d}\ln{\Omega}}{{\rm d}\ln{R}}\right)^{2} = 2\sigma_{SB}T_\text{eff}(R)^4 \ .
\end{equation}
Here, $\sigma_{SB}$ is the Stefan-Boltzmann constant and in the optically-thick limit, the central disk temperature $T_0$ and the photospheric temperature $T_\text{eff}$ are related as $\left(\dfrac{T_0}{T_\text{eff}}\right)^{4} \simeq \dfrac{3}{4}\tau$, for (vertical) optical length $\tau$ which in our case we take to be arising from Thomson opacity ($\kappa_e = \dfrac{\sigma_T}{m_p}$, $\sigma_T$ being the Thomson cross-section and $m_p$ being the proton mass), i.e. $\tau(R) = \dfrac{1}{2} \dfrac{\sigma_T}{m_p} \Sigma(R)$.\footnote{A similar exercise can be repeated for other opacity sources: e.g. bound-free opacity or free-free opacity. Our goal here is to derive some simple scaling relations, so in this paper we shall only limit ourselves to electron-scattering opacity for illustrative purposes.} 

Inserting $T_{\rm eff}^{4} \propto T_{0}^{4}/\tau \propto T_{\rm 0}^{4}/\Sigma$ in equation (\ref{eq:heat_balance}) gives $T_{0}^{4} \propto \nu \Sigma^{2} \Omega^{2}$ or $\propto \dot{M} \Sigma \Omega^{2}$ with equation (\ref{eq:visc_mdot}), which in equation (\ref{eq:rad.ve}) gives $v_{e}^{2} \approx (H\,\Omega)^{2} \propto \dot{M} \Sigma \Omega^{2} / (\Sigma/H)$ or $H \propto \dot{M}$. In other words, the result that $H$ is constant with radius (up to the order-unity constant ${\rm d}\ln{\Omega}/{\rm d}\ln{R}$) and independent of $\alpha$ follows, with (retaining the various constants above): 
\begin{equation}
    H(R) = \frac{\sigma_T\dot{M}}{4\pi\,c\,m_p}\left|\frac{{\rm d}\ln{\Omega}}{{\rm d}\ln{R}}\right| \propto \dot{M} = \text{const} \ .
\end{equation}
Defining the dimensionless Eddington accretion ratio $\dot{m} \equiv 0.1\,\dot{M}\,c^{2} / L_{\rm Edd}$, scaled mass $m_{8} \equiv M_{c} / 10^{8} M_{\odot}$, $\alpha_{0.1} \equiv \alpha/0.1$, and scaled radius $r \equiv R/R_{\rm S}$, this means
\begin{equation}
\frac{H(R)}{R} \sim \frac{10\, \dot{m}}{r} \ .
\end{equation}

This implies that at large radii, radiation pressure supported disks look increasingly thinner: $H/R \propto R^{-1}$; in fact, they have the lowest aspect ratios beyond a few thousand gravitational radii of any disk model we discuss. With this, it is straightforward to derive the relation between the surface density profile $\Sigma(R)$ and the rotational profile $\Omega(R)$, using equation (\ref{eq:visc_mdot}) and the definition of $\nu$: $\dot{M} \propto \nu \Sigma \propto \alpha v_{e} H \Sigma \propto \alpha H^{2} \Omega \Sigma \propto \alpha \dot{M}^{2} \Omega \Sigma$, or specifically: 
\begin{equation}
    \Sigma(R) = \frac{8\pi\,c^2 m_p^2}{\alpha \sigma_T^2\dot{M}\Omega(R)}\left|\frac{{\rm d}\ln{\Omega}}{{\rm d}\ln{R}}\right|^{-3}\ .
\end{equation}

So, we see, $\Sigma(R)\propto \Omega(R)^{-1}$. Within the SMBH radius of influence, the rotation is largely Keplerian, dominated by the SMBH potential, so $\Omega(R) \approx \sqrt{G M_{c} / R^{3}}$ and 
\begin{equation}
    \Sigma(R) \rightarrow \frac{64\pi}{27} \frac{c^2 m_p^2}{\alpha\sigma_T^2\dot{M}}\sqrt{\frac{R^3}{GM_c}} \propto R^{3/2}\ .
\end{equation}

Beyond the radius of influence of the central SMBH, given as,
\begin{equation}
    R_\text{ROI} = \left(\frac{189}{256\pi^{2}} \frac{\alpha\sigma_T^2G^{1/2}\dot{M}}{c^2m_p^{2}}\right)^{2/7}M_c^{3/7}
\end{equation}
the self-gravity of the disk starts to dominate the global enclosed mass and disk dynamics. More explicitly, we need to include the disk mass in computing the rotational profile $\Omega(R)^2 = \dfrac{G}{r^3}\left(M_c+M_d(R)\right)$. At large radii, where $R> R_{ROI}$, and $M_d(R)>M_c$, we can approximate $\Omega(R)^2\simeq \dfrac{2\pi G}{R^3}\int_0^R{\rm d}R'\text{ }R'\Sigma(R')$, to obtain the surface density profile at large radii,
\begin{equation}
    \Sigma(R) = 36\left(\frac{7\pi c^4m_p^4R}{6 G\alpha^2\sigma_T^4\dot{M}^2}\right)^{1/3} \propto R^{1/3}
\end{equation}
It follows, then, $\Omega(R)\propto R^{-1/3}$.

With these, we can also calculate ${\rm d} \Gamma^{\rm tot}/{\rm d} \ln{R}$ from equation (\ref{eq:effrts}), up to the uncertainty in $m_{1} f_{1} f_{t}/m_{t}$ (which depends on the objects and processes of interest). It is convenient to write this in the normalized units: 
\begin{equation}
\tilde{\Gamma}^{\rm tot} \equiv  \frac{ {{\rm d} \Gamma^{\rm tot}} / {{\rm d} \ln{R}} }{{\rm Gyr^{-1}} (m_{1}/m_{t}) f_{1} f_{t} }
\end{equation}
\begin{equation}
\tilde{\Gamma}^{\rm tot}  \sim 
    \begin{cases}
    10^{19.7} m_{8} \dot{m}^{-6}  r_{4}^{19/2} & ;R\leq R_\text{ROI}\\
    10^{12.7} m_{8}^{-2/3} \alpha_{0.1}^{5/3} \dot{m}^{-13/3}  r_{4}^{11/3}  & ;R> R_\text{ROI}
    \end{cases}
\end{equation}
where we have also defined $r_{4} \equiv r / 10^{4} = R/10^{4} R_{\rm S}$.

We must note here that our calculation in this section assumes a radiation pressure supported disk where the radiation pressure comes from the accretion luminosity. Therefore, the scalings we derive here are different from those in \cite{tqm}, which is a popularly invoked disk model that describes a radiation pressure supported disk; however in their disks, the vertical support comes from radiation pressure on dust grains from star formation and stellar feedback which aid to self-adjust the Toomr\'{e} $Q$ parameter in these disks at $Q\sim1$.  So, the radiation support in their case is equivalent to turbulent pressure support, and their disks are, thus, marginally-stable gravito-turbulent disks. We discuss self-gravitating turbulent pressure supported disks in \S\ref{sec:turb_disk}.

\subsection{Thermal Gas Pressure-Dominated Disks}

Here, we consider similar set-up as the previous section except that now we take the thermal gas pressure (following the ideal gas equation of state, so the sound speed $c_s^2(R) \sim \dfrac{k_B}{ m_p}T_0(R)$ in the midplane where $k_B$ is the Boltzmann constant) to be providing the dominant vertical support. Angular momentum transport through the disk is still assumed to be affected by effective viscous torques, following an $\alpha$-parameterisation for specifying the viscosity. This implies, in conjunction with equations (\ref{eq:mdot}), (\ref{eq:visc_mdot}) and (\ref{eq:scaleht}), that $v_e(R)^2 \simeq \alpha c_s(R)^2$.

The temperature profile of the disk, as before, is set by the energetic balance between heating via viscous dissipation and cooling via thermal radiation (equation (\ref{eq:heat_balance})). Therefore (again taking $\tau \rightarrow \sigma_{T} \Sigma/2m_{p}$) we obtain:
\begin{equation}
    H(R)\simeq\frac{v_e(R)}{\Omega(R)} 
    = \left(\left| \frac{{\rm d} \ln \Omega}{{\rm d} \ln R} \right|^{2}  \frac{3 \sigma_{T} k_{B}^{4} \dot{M} \Sigma(R) }{16 m_{p}^{5} \sigma_{\rm SB} \Omega^{6} } \right)^{1/8}
\end{equation}
And combined with equation (\ref{eq:visc_mdot}) and the definition of $\nu$: 
\begin{equation}
\begin{split}
\Sigma(R) \simeq \frac{\sigma_{\rm SB}^{1/5} m_{p} \dot{M}^{3/5} \Omega^{2/5} }{(3\sigma_{T})^{1/5} (\alpha \pi k_{B})^{4/5} } \left|  \frac{{\rm d} \ln \Omega}{{\rm d} \ln R}\right|^{-6/5}
\end{split}
\end{equation}

Such a disk dominated by gas pressure and electron-scattering opacity forms the central region of popular disk models such as the classic Shakura-Sunyaev \cite{SS73} or Novikov-Thorne model. However, we draw attention to an important distinction here. Most of these models (such as the \cite{SS73} model) assume a Keplerian potential. But, as we have stressed before, it becomes essential to account for self-gravity effects outside the radius of influence of the black hole (which are typically not included in the scaling expressions popularly cited). For our assumptions, this radius corresponds to
\begin{equation}
    R_\text{ROI} \sim 6000\, R_{\rm S}\,\frac{\alpha_{0.1}^{4/7}}{m_{8}^{6/7}\,\dot{m}^{3/7}} \ .
\end{equation}

Taking these corrections into account, the scalings for $\Omega(R)$ and $\Sigma(R)$ become: 
\begin{equation}
    \Sigma(R)\propto\Omega(R)^{2/5}\propto
    \begin{cases}
    R^{-3/5} & ;R\leq R_\text{ROI}\\
    R^{-1/4} & ;R> R_\text{ROI}
    \end{cases}
\end{equation}
and the scaling for $H(R)/R$ becomes: 
\begin{equation}
    \frac{H(R)}{R}\sim 
    \begin{cases}
    0.003\, \dot{m}^{1/5} (\alpha_{0.1} m_{8})^{-1/10} r^{1/20} & ;R\leq R_\text{ROI}\\
    0.6\, \dot{m}^{-1/16} \alpha_{0.1}^{1/4} m_{8}^{-5/8} r^{-9/16}  & ;R> R_\text{ROI}
    \end{cases}
\end{equation}
So there is a qualitative shift beyond the radius of influence of the SMBH; the self-gravity of the gas actually results in lower disk aspect-ratios as compared to a purely Keplerian potential. In fact, the aspect ratio ceases to gradually increase outwards, and instead shows a relatively steeper decrease with radius.

Combining with equation (\ref{eq:effrts}) gives:
\begin{equation}
\tilde{\Gamma}^{\rm tot}  \sim 
    \begin{cases}
    10^{14.3} m_{8}^{9/5} \alpha_{0.1}^{4/5} \dot{m}^{2/5}  r_{4}^{11/10} & ;R\leq R_\text{ROI}\\
    10^{14.1} m_{8}^{9/4} \alpha_{0.1}^{1/2} \dot{m}^{5/8}  r_{4}^{13/8}  & ;R> R_\text{ROI}
    \end{cases}
\end{equation}
for the implied event rates.

\subsection{Self-Gravitating Turbulence-Supported Disks}\label{sec:turb_disk}

In ``marginally stable'' self-gravitating disk or ``gravito-turbulent'' models, the effective turbulent Toomr\'{e} stability parameter $Q_\text{turb}$ is assumed to self-consistently regulate via highly super-sonic turbulence in some stable feedback loop. Essentially, if $Q_{\rm turb}$ becomes too high, the self-gravity becomes unimportant but the turbulence dissipates on its own crossing time via shocks and/or viscous damping, so $Q_{\rm turb}$ decreases, while if $Q_{\rm turb}$ becomes too low, gravitational instability leads to collapse and increased turbulence or bulk motions which ``pump up'' $Q_{\rm turb}$. In steady-state, this is assumed to maintain  
\begin{equation}\label{eq:Toomre}
    Q_\text{turb}=\frac{\delta v\text{ }\kappa_{\rm epi}}{\pi G \Sigma}\sim1
\end{equation} 
with $\kappa_{\rm epi}^{2} \equiv 2\,\Omega^{2} (2 + {\rm d}\ln\Omega/{\rm d}\ln{R})$ the epicyclic frequency, 
and the disk is therefore (by assumption) supersonically and super-Alfvenically turbulent with the turbulent eddies (Reynolds stresses) dominating the angular momentum transport through the disk. 
Note that this expectation is robust to the detailed physics driving the turbulence, whether self-gravity alone \citep{gammie:2001.cooling.in.keplerian.disks,SG2003} or feedback (supernovae, winds, radiation) injected by stars or compact objects embedded within the disk, as in some models \citep{tqm}. 
Since the turbulent motions $\delta v \sim v_{e}$ are the dominant pressure this produces a scale-height
\begin{equation}
    H(R) \simeq \frac{\pi G\Sigma(R)}{\Omega^2(R)}\ .
\end{equation}
Here we neglect the details of anisotropy in the turbulence, but per our discussion above this does not change the qualitative predictions of these models.

Using the condition (\ref{eq:Toomre}) with equation (\ref{eq:RMstress}) describing mass transport via Reynolds stresses, then interior to the ROI (where the potential is close to Keplerian) we obtain the surface density profile in the disk
\begin{equation}
    \Sigma(R) \simeq \frac{1}{\pi}\left(\frac{\dot{M}}{3 G^2}\right)^{1/3}\Omega(R) 
\end{equation}
i.e., $\Sigma(R) \propto \Omega(R)$, and $H(R) \approx (G \dot{M}/3)^{1/3} \Omega^{-1}$. The potential is primarily Keplerian ($\Omega \approx \sqrt{G M_{c}/R^{3}}$) within the region of interest here. Therefore, in this region,
\begin{equation}
 \frac{H(R)}{R} \sim \frac{\dot{M}^{1/3} R^{1/2}}{3^{1/3} G^{1/6} M_{c}^{1/2}} \sim 10^{-4} m_{8}^{1/3} \dot{m}^{1/3} r^{1/2}
\end{equation}
 and the disk get thicker in the outer regions. 
With equation (\ref{eq:effrts}) we have:
\begin{equation}
\tilde{\Gamma}^{\rm tot}  \sim 10^{10.7} m_{8}^{-5/3} \dot{m}^{-2/3} r_{4}^{-5/2}\ .
\end{equation}
 
Outside the radius of influence of the SMBH when the mass of the disk starts to dominate gravitationally ($M_{d} \gtrsim M_{c}$), $Q_{\rm turb} \approx 1$ requires $H \approx R$, so the aspect ratio $H/R$ saturates at unity (with $\Sigma \rightarrow 2\,\dot{M}^{2/3}/(\pi G^{1/3} R)$, so $\Omega \propto R^{-1}$). 

\subsection{Magnetic Pressure-Supported Disks}

A strongly magnetized disk with non-negligible toroidal (or radial) fields in the midplane ($\beta_{\rm thermal} \equiv P_{\rm thermal,\,gas}/P_{\rm B} \lesssim $ 1), and trans-Alfv\'{e}nic turbulence (as argued for in \citet{mag_analytic} and motivated by recent zoom-in simulations of accretion disks around SMBH by self-consistently following gas in cosmological simulations down to within the accretion disk in quasar-like systems \citep{Forgd,Forgd2,hopkins:superzoom.agn.disks.to.isco.with.gizmo.rad.thermochemical.properties.nlte.multiphase.resolution.studies,kaaz:2024.hamr.forged.fire.zoom.to.grmhd.magnetized.disks} or low-luminosity AGN \citep{guo:2024.fluxfrozen.disks.lowmdot.ellipticals} or IMBHs and low-mass SMBHs in star clusters and proto-galaxies \citep{yanlongGMC,shi:2024.seed.to.smbh.case.study.subcluster.merging.pairing.fluxfrozen.disk} or circum-binary-SMBH disks \citep{wang:2025.mad.magnetized.circumbinary.disks.strong.bfields.large.cavities.scale.of.binary}), is characterized by a much puffier geometry that flares more strongly than any of the previous models in the outer regions, approaching $H\sim R$. We quote here the relevant scalings (see \cite{mag_analytic} and \cite{Pariev} for derivation and discussion),
\begin{equation}
    \frac{H(R)}{R}\simeq \left(\frac{R}{R_\text{ROI}}\right)^{1/6} \sim 0.1\,m_{8}^{1/12} r^{1/6}
\end{equation}
\begin{equation}
    \Sigma(R) \simeq \frac{\dot{M}}{2\pi\sqrt{GM_cR_\text{ROI}}}\left(\frac{R}{R_\text{ROI}}\right)^{-5/6}
\end{equation}
where the radius of influence of the central SMBH is fixed as $R_\text{ROI} \sim 5\text{ pc }\sqrt{M_c/(10^7\text{ M}_\odot)}$ as in \cite{mag_analytic} guided by numerical simulations such that the accretion onto the SMBH is ultimately driven by gravitational capture of gas moving roughly at the dispersion velocity $\sigma_\text{gal}$ in the galaxy which can be related to the SMBH mass $M_c$ using the observed $M-\sigma$ relation \citep{KormendyHo}. Similar scalings apply to the inner regions of magnetically-arrested disks (MAD) in simulations, even if the details of the magnetic geometry differ between models \citep[see][]{liska:2020.grmhd.sims.toroidal.becomes.poloidal,kaaz:2022.grmhd.sims.misaligned.acc.disks.spin,kaaz:2024.hamr.forged.fire.zoom.to.grmhd.magnetized.disks}.

With equation (\ref{eq:effrts}) this gives:
\begin{equation}
\tilde{\Gamma}^{\rm tot}  \sim 10^{-2} m_{8}^{1/3} \dot{m}^{2} r_{4}^{1/6}\ .
\end{equation}

\subsection{More General Models}\label{sec:general.models}

Of course, any realistic astrophysical accretion disk model will include some combination of the various models mentioned above. For example, the widely-invoked thin $\alpha$-disk Shakura-Sunyaev model \citep{SS73} has an inner region that is dominated by radiation pressure and electron-scattering opacity, a middle region that is thermal gas pressure dominated with Thomson scattering still the primary opacity source, and an outer region where the dominant opacity source changes to line opacity. The Sirko-Goodman model \citep{SG2003} also has an inner radiation pressure dominated region, which transitions to an outer region where the gravito-turbulent stresses self-regulate the disk to maintain $Q_\text{turb} \sim 1$. The Thompson-Quataert-Murray model \citep{tqm} connects an inner (radiation pressure dominated, stable) $\alpha$-disk to an outer disk where radiation pressure from star formation and supernovae is assumed to regulate the disk at marginal stability $Q\simeq1$. The Begelman-Pringle model \citep{BP2007} assumes an $\alpha$-disk model that is magnetically-dominated. And the simulations in \citep{kaaz:2024.hamr.forged.fire.zoom.to.grmhd.magnetized.disks,hopkins:superzoom.agn.disks.to.isco.with.gizmo.rad.thermochemical.properties.nlte.multiphase.resolution.studies} with super-Eddington accretion rates transition to an inner (near-horizon) region within tens of $R_{\rm S}$ where magnetic and radiation pressure are comparable. And of course there will be deviations in detail from the highly simplified analytic similarity solutions we use for the scalings above.

Moreover, there are some regions of this parameter space that are already ruled-out by observations or basic physical consistency requirements. For example, it becomes more and more unlikely that disks would be radiation-pressure-dominated at large radii: beyond $\gg 100\,R_{\rm S}$, this is easily ruled-out, as the required radiation energy densities at large radii would be so large that the luminosities of the AGN would necessarily exceed the Eddington luminosity by orders of magnitude, which is not observed. Likewise, beyond a radius $\sim 100-1000\,R_{\rm S}$, thermal-pressure dominated models become gravitationally strongly unstable (Toomre $Q \ll 1$). Conversely, at small radii $\lesssim 10^{3}\,R_{\rm S}$, the turbulent pressure models predict that heating from the dissipation of turbulence itself would necessarily lead to thermal and/or radiation pressures larger than the turbulent pressure. Disks with large magnetic pressure, however, are allowed over the entire range of radii plotted.

Regardless, even in these more detailed hybrid models and numerical simulations, the simple scalings above an in Figure~\ref{fig:hor} still provide a reasonable (order-unity accurate) approximation of $H/R$ and $\Sigma(R)$ at each radius, provided that one chooses the model corresponding to what dominates the pressure \textit{at each given radius}. In other words, to lowest order, these models more or less interpolate between the different lines in Figure~\ref{fig:hor} corresponding to the dominant source of pressure in the model at each $R$. The relevant values of $H/R$ for a specific physical question -- like, for example, computing accretion rates onto stars or BHs embedded in the AGN disk, or binary pairing/hardening rates -- depend on the source of pressure at the radii of interest. For many of the processes in \S\ref{sec:intro}, this will be the outer disk at radii $\gg 1000\,R_{\rm S}$.

\section{Capture Rates}\label{sec:rates}

\begin{figure}
    \centering
    \includegraphics[width=0.98\columnwidth]{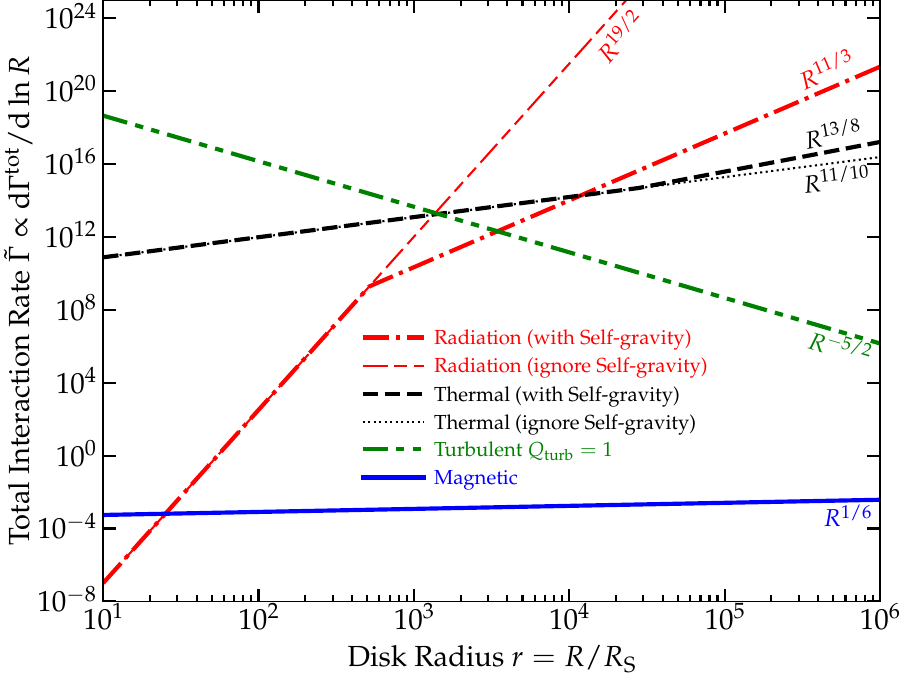}\vspace{-0.3cm}
    \caption{The total two-body capture/interaction/accretion/merger/friction/migration rate, as a function of radius for different disk models (as Fig.~\ref{fig:hor}, same model parameters assumed). The vertical axis has arbitrary units because the absolute number/mass of interacting objects depends on the specific model assumed, but note the extremely large range in orders of magnitude. Disks with multiple comparable forms of pressure will essentially trace the lowest curve here of the different pressure terms assumed, so ``adding pressure'' strongly decreases the rates. Regardless of details, rates are extremely sensitive to the pressure source at each $R$. Historical models for these processes have generally assumed something between the thermal and turbulent lines here, and largely neglect magnetic pressure, but this shows that magnetic fields can lower the predicted rates by $\sim 20-30$ orders of magnitude.}
    \label{fig:cap_rts}
\end{figure}

In Figure \ref{fig:cap_rts}, we illustrate how the assumption different disk pressure or physics qualitatively modifies the expected binary capture/merger/accretion/harding/dynamical friction/migration rates in these disks. From the (more general) expression in equation (\ref{eq:netrate}), note that the total interaction rate per radial bin in the disk scales as
\begin{equation}
    \frac{{\rm d}\Gamma^\text{tot}(R)}{{\rm d}\ln{R}} \propto \Sigma^2(R) \frac{\Omega(R)^{-3}}{R^2} \left(\frac{H(R)}{R}\right)^{-4} \ .
\end{equation}

We plot this factor as a function of the radial position in the disk for the different disk models in Figure~\ref{fig:cap_rts}. It is noteworthy how different disk physics can result in several tens of orders-of-magnitude difference in the predicted rates. For all the disk models, the mass per annulus of the disk increases outwards (so the number of available bodies to gravitationally interact with each other increases) but the disks may become thinner or thicker (thereby spreading the available bodies over a larger volume), so there can be a competing effect between the two which determines how the capture rate increases/decreases with radius. In radiation pressure dominated disks, as the aspect ratio becomes thinner towards larger radii, and the surface density increases, we observe a steep rise in the number of gravitational capture events predicted in the outer regions (though as noted above, it is probably already ruled-out that disks are radiation pressure-dominated out to such large radii). 
Again, we make a distinction between the self-gravity effects that become important outside the SMBH radius of influence, over the strictly Keplerian case, and modify the surface density profile (and thus the rotation curve in the disk) accordingly to recover a comparatively much less steep rise in the expected rate of capture events in the outer radii. The gravito-turbulent disk that is supported by supersonic turbulence against collapse sees both a surface density drop as well as aspect ratio increase with radius, so the capture rate decays strongly with increasing radius. The strongly-magnetized disks show the most gradual trend with radius, and their high aspect ratios put the expected rate magnitudes many orders of magnitude below other disk models at almost all radii.

Again, it is important to mention here that in all these cases, we have neglected the radial dependence of how the factors $f_1, f_t$ that encapsulate the relative occurrence of the interacting species in different regions of the disk. These terms would depend strongly on the species being considered. As an example, if the target species were stars in a marginally stable self-gravitating disk, we expect $f_t$ to be small within the radius where the (turbulent) Toomr\'{e} $Q$ parameter is about unity; and to increase beyond the radius where it falls below $Q\lesssim1$ allowing efficient fragmentation (this radius is roughly the radius of influence of the SMBH). Still, the dependence on disk geometry is so much stronger that for qualitative purposes these distinctions should not represent the dominant difference between model predictions.

\section{Implications}\label{sec:implications}

\subsection{Gravitational wave (and electromagnetic) merger/event rates for LIGO/Virgo/KAGRA}

A number of studies have suggested constraints for the rates of BH-BH mergers occurring in AGN disks, which may be detectable as gravitational wave sources \citep[e.g.,][]{EvolutionOfCompactObjsInDisks,Bartos2018,Stone2017,McKernan2018}. The estimate, for their choice of disk parameters, tends to be around $1-10$ Gpc$^{-3}$ yr$^{-1}$, roughly consistent with the local BH merger rate inferred from the latest catalogue of LIGO/Virgo/KAGRA detections \citep{LVK2023}. Per \S\ref{sec:intro}, similar assumptions have been used to postulate that other mergers (of e.g.\ stars, white dwarfs, neutron stars, etc.) in circum-SMBH disks could explain different flares or transient behaviors of AGN, or induce rapid rates of tidal disruption events of stars interacting with the central SMBH. However, all of these analyses use some version of the \cite{SS73} or \cite{tqm} ``razor-thin-disk'' models (similar $H/R$ at the relevant radii to the thermal-pressure-dominated similarity model in Figure~\ref{fig:hor}). 
As shown in the previous section, assuming the disk is so thin greatly overpredicts merger rates compared to a thicker disk models, either those with significant magnetic pressure and/or stronger turbulence or $Q \gtrsim 1$. 

For instance, notice that the disks with significant magnetic pressure in Figure \ref{fig:cap_rts} predict rates \textit{tens of orders of magnitude} lower than the fiducial disk models used in these prior rate calculations. And there are a number of recent theoretical and observational arguments favoring significant magnetic pressure at these radii (the outer regions of the disk at $\gg 1000\,R_{\rm S})$. For example, recent numerical simulations, \cite{Forgd,Forgd2,hopkins:superzoom.agn.disks.to.isco.with.gizmo.rad.thermochemical.properties.nlte.multiphase.resolution.studies,kaaz:2024.hamr.forged.fire.zoom.to.grmhd.magnetized.disks} zooming-in from cosmological scales to $\lesssim 100\,R_{\rm S}$ to study the self-consistently formed AGN disks find they are highly (super-sonically, trans-Alfv\'{e}nically) turbulent with magnetic pressure dominant ($\beta_{\rm thermal} \sim 10^{-6}-10^{-2}$ in the mid-plane) and aspect ratios as large as $H/R \sim 0.3-0.5$. These findings are consistent with idealized experiments using different initial conditions and numerical methods \citep{johansen.levin:2008.high.mdot.magnetized.disks,gaburov:2012.public.moving.mesh.code,Squire_2025,guo2025idealizedglobalmodelsaccretion,tomar.hopkins:2025.lagrangian.vs.eulierian.methods.toroidally.magnetized.disks.guo.test.problem.would.see.collapse.even.low.res.in.mfm.mfv}, multi-physics simulations of accretion onto IMBHs and low-mass SMBHs in dense star clusters or galactic nuclei \citep{yanlongGMC,shi:2024.seed.to.smbh.case.study.subcluster.merging.pairing.fluxfrozen.disk}, circum-binary AGN disks forming from interstellar-medium scales \citep{wang:2025.mad.magnetized.circumbinary.disks.strong.bfields.large.cavities.scale.of.binary}, zoom-in simulations from circumgalactic medium to accretion disk scales in low-luminosity AGN \citep{guo:2024.fluxfrozen.disks.lowmdot.ellipticals} and simulations of accretion onto SMBH seeds formed from direct collapse \citep{Schlosman}. And these disks appear to agree much more closely with constraints at these radii on disk masses and aspect ratios from resolved kinematics of broad-line regions, microlensing, molecular masers, and direct imaging from interferometry, as well as constraints on the outer disk magnetic fields from Zeeman splitting \citep{Pariev,hopkins:multiphase.mag.dom.disks,hopkins:2025.maser.evidence.for.mag.disks}. But these produce qualitatively different predictions from the thin-disk profiles most analyses assume, including the many aforementioned studies estimating GW merger rates in AGN environments. 

Hence, we conclude the merger rates quoted above may be highly optimistic.

\subsection{Accretion and Fueling of Compact Objects or Stars in Disks}

As we note above, the same arguments apply to any other gravitational capture processes in AGN accretion disks. That includes accretion onto compact objects (stellar or intermediate-mass BHs, neutron stars, white dwarfs) or stars within the disk, and their implied consequences for explaining phenomena like transients or features in AGN spectra or supermassive stars or ``immortal'' (rapidly-accreting) massive stars (\S\ref{sec:intro}). All of these prior calculations generally assumed standard, geometrically very thin (often thermal-pressure-dominated) $\alpha$-disk type models. So the same orders-of-magnitude caveats apply to these calculations, and it is similarly likely that the prior calculations essentially define the ``upper limit'' or most-optimistic case for accretion processes onto stars/compact objects within the disk (with e.g.\ other disk models featuring stronger turbulence and/or magnetic support predicting many orders-of-magnitude lower rates).

Note for example, if we take the ``target'' species to be gas, and multiply the capture rate $\Gamma$ (equation (\ref{eq:gamma})) by $m_{t}$ we just obtain the usual accretion rate for an embedded mass $m_{1}$, which gives the mass-doubling timescale $t^{(1)}_{\rm acc} \sim m_{1} / \dot{m}_{1} \sim m_{1} / (m_{t} \Gamma) \sim 4 \Omega^{4} H^{6} / (G^{2} \dot{M} m_{1})$. Considering e.g.\ a typical stellar-mass BH with $m_{1} \sim 10\,M_{\odot}$, in the outer radii ($r \gtrsim 10^{4}$) of the radiation or thermal-dominated cases above, this gives an accretion time orders-of-magnitude shorter than a single orbital time -- i.e.\ embedded stars and BHs would grow incredibly rapidly in the outer disk. For the turbulent ($Q_{\rm turb} \sim 1$) case the accretion time is longer than orbital in the outer disk but still shorter than the viscous timescale of the disk ($\sim 2\pi R^{2}/\nu$) at $r \lesssim 10^{4} (m_{8} \dot{m})^{-2}$. While for the magnetic case, $t^{(1)}_{\rm acc}$ would exceed $\gtrsim 10^{13}\,$yr, much longer than the Hubble time.

\subsection{Gap-Opening in AGN disks}

Briefly, consider the physics of gap-opening in AGN disks, which as we noted above scales differently with $H/R$. It is well-known that there are two necessary conditions for gap opening (and related processes like gap-fed accretion, Type-II migration, escape of electromagentic radiation from gaps without scattering/reprocessing by the disk, etc.) to occur in any form \citep{takeuchi:1996.gap.formation.disk.criteria.review}. Heuristically, the first is that the Hill radius $R_{H} \approx q^{1/3}\,R$ (where $q \equiv m_{1}/M_{c}$) must be larger than $H$ (or else gas will simply flow around the test mass and accretion/capture onto it will be Bondi-Hoyle-Lyttleton-like and follow the scalings in \S\ref{sec:Formalism}), i.e.\ $q \gtrsim (H/R)^{3}$. The second is that the torque from the planet must exceed the viscous stress from the disk (or equivalently that the gap-opening rate must exceed the gap-refilling rate from accretion), which can be written $q \gtrsim (H/R^{2})\,(\nu/\Omega)^{1/2} \sim \alpha^{1/2} (v_{e}/\Omega R)^{2} \sim (H/R)^{2}$. 
For most of the parameter space considered here, the latter is the more demanding criterion, and (since $\alpha\sim \mathcal{O}(1)$ as defined here for the turbulent and magnetic models) implies that gap opening can only occur if the test mass $m_{1}$ exceeds a critical 
\begin{equation}
m_{1} \gtrsim m_{1}^{\rm crit,\,gap} \sim (H/R)^{2} M_{c}\ .
\end{equation}
In either case ($(H/R)^{3}$ or $(H/R)^{2}$), this is still quite sensitive to $H/R$, just not as extreme as the prior scaling for binary interactions or accretion. 

From Figure \ref{fig:hor}, we see this still varies dramatically (by up to $\sim 10$ orders-of-magnitude) depending on the disk physics. For the fiducial models in \S\ref{sec:models}, this gives 
\begin{equation}
    \frac{m_{1}^{\rm crit,\,gap}}{{\rm M_{\odot}}} \sim 
    \begin{cases}
    10^{9.5} m_{8} \dot{m}^{2} r^{-2} & ;\,{\rm (radiation)}\\
    300 m_{8}^{4/5} \dot{m}^{2/5} r^{1/10} & ;\,{\rm (thermal)}\\
    2\,m_{8}^{5/3} \dot{m}^{2/3} r & ;\,{\rm (turbulent)}\\
    10^{6} m_{8}^{7/6} r^{1/3} & ;\,{\rm (magnetic)}\\
    \end{cases}
\end{equation}
(within the Keplerian radii for each). 
For e.g.\ the radiation or thermal cases, this implies that for a lower-mass SMBH (say $\sim 10^{6}\,M_{\odot}$) in a typical luminous state ($\dot{m} \sim 0.1$), a standard stellar-mass BH with mass of just $m_{1} \sim 10\,{\rm M}_{\odot}$ would be sufficient to drive gap opening (and Type-II migration), placed almost anywhere in the disk (at which point it will quickly reach a mass  $m_1 \rightarrow 2\pi R\Sigma(R)R_H$, and then capture the entire accretion flow onto itself preventing flow through the gap, i.e.\ growing at $\dot{m_1} \rightarrow \dot{M}$). This presents a challenge, as such efficient gap-opening would imply that just a few stars or BHs around the accretion disk should be sufficient to completely shut down accretion onto the SMBH/inner disk (making it difficult to understand how any luminous lower-mass AGN can exist; see \cite{goodman:qso.disk.selfgrav}). For the turbulent case the same BH would cause gap opening interior to $r \lesssim 6\times10^{4}$ or $\lesssim 0.006\,$pc, and ``supermassive stars'' or IMBHs would drive gap opening everywhere even for more massive SMBHs. For the magnetic case, gap opening at almost any SMBH mass or radius of interest requires $m_{1} \gtrsim 0.1\,M_{c}$ -- i.e.\ a massive/major SMBH-SMBH merger, well beyond the ``test mass'' (or stellar-mass BH or star) limit for $m_{1}$. 

Moreover, since the sign of these effects with $H/R$ is similar to those of accretion reviewed above, it is much more difficult for any object to grow ``up to'' a mass sufficient for gap-opening at larger $H/R$. In other words, in the thicker models, objects sufficient to cause gap opening cannot grow to this mass in-situ but must be accreted or merge into the SMBH parent system, while in the thinnest models it is almost ensured that any test particle placed into the disk will rapidly grow to a mass sufficient to cause gap opening. Therefore, while  traditional gap-opening and migration pictures -- largely based on planetary dynamics in circumstellar disks -- may still apply in the models closer to ``razor thin'' disks, we see that disk thickness and pressure support plays a key role here and can strongly suppress this process. 

\section{Conclusion}\label{sec:conclusion}

In this paper, we have performed a simple analytical calculation to highlight the importance of considering the dominant physics which governs the geometric structure of accretion disks in predicting relative gravitational capture rates. Any process that is mediated by the gravitational cross-section scales very strongly with the aspect ratio of these disks. Since different disk models can differ by several orders of magnitude in the $H/R$ profile at any given location in the disk, this translates to tens of orders of magnitude difference in the rates predicted for events like GW-mergers, Bondi-Hoyle accretion onto compact objects embedded in the disk, etc. In the past, most studies have assumed the outer disks to be very thin (with aspect ratios $\lesssim 0.01$). But more recent theoretical arguments, numerical simulations forming AGN disks from more self-consistent interstellar medium conditions, and observational constraints have argued these disks may in fact have thicker, ``puffy'' geometries. This would imply that the expected rates of binary interactions, mergers, accretion onto stellar-mass objects or IMBHs, gap-opening, and other gravitational processes in AGN disks may have been greatly overestimated by past analyses that assumed older thin disk models.

\begin{acknowledgments}
PFH and YT were supported by a Simons Investigator grant. We thank the anonymous referee for a number of helpful comments.
\end{acknowledgments}

\bibliographystyle{apsrev4-2}
\bibliography{ms_local}

\end{document}